\documentclass{article}
\usepackage{spconf,amsmath,graphicx}
\usepackage{multirow}
\usepackage{color}
\usepackage{booktabs}
\usepackage{amsfonts}
\usepackage{subfigure}
\usepackage{algorithm}
\usepackage{algorithmic}
\usepackage{epsfig,epstopdf}

\title{RACORN-K: Risk-Aversion Pattern Matching-based Portfolio Selection}
\name{Yang Wang$^1$, Dong Wang$^{1*}$, Yaodong Wang$^1$, You Zhang$^2$\thanks{This work was supported in
part by the National Natural Science Foundation of China under Projects 61371136 and 61633013. 
Corresponding Author: Dong Wang.}}
\address{$^1$Center for Speech and Language Technologies, Tsinghua University, China\\ $^2$ Derivatives China, China\\
{\small \tt \{wangyang,wangyd\}@cslt.riit.tsinghua.edu.cn}\\
{\small \tt wangdong99@mails.tsinghua.edu.cn, zy@derivatives-china.com}
}
\begin{document}
%
\maketitle
\begin{abstract}

Portfolio selection is the central task for assets management, but it turns out to
be very challenging. Methods based on pattern matching, particularly the
CORN-K algorithm, have achieved promising performance on several stock markets.
A key shortage of the existing pattern matching methods, however, is that
the risk is largely ignored when optimizing portfolios, which may lead to
unreliable profits, particularly in volatile markets.
We present a risk-aversion CORN-K algorithm, RACORN-K, that
penalizes risk when searching for optimal portfolios. Experiments on
four datasets (DJIA, MSCI, SP500(N), HSI) demonstrate that the new
algorithm can deliver notable and reliable improvements in terms of return,
Sharp ratio and maximum drawdown, especially on volatile markets.

\end{abstract}
\section{Introduction}
\label{sec:intro}

Portfolio selection has gained much interest for its theoretical importance and practical value.
It aims at optimizing the assets allocation so that higher returns can be obtained while taking less risk.
According to the assumptions of the financial signal, existing portfolio selection strategies can be classified into
three categories~\cite{li2014online}: follow-the-winner~\cite{cover1991universal, helmbold1998line, agarwal2006algorithms, singer1997switching}, follow-the-loser~\cite{borodin2004can, li2012pamr, li2013confidence, li2015moving}, and
pattern matching~\cite{gyorfi2006nonparametric,gyorfi2008nonparametric,li2011corn}. The first two categories
heavily rely on the trend of the market, thus may lead to a huge loss if the trend is not as assumed~\cite{jegadeesh1990evidence}.
The pattern matching approach, in contrast, relies on a more practical assumption that \emph{patterns will reoccur},
hence more practically applicable.

A typical pattern matching algorithm involves two stages: similar period retrieval and portfolio optimization.
Most of the existing researches focus on the first stage, in particular how to measure the similarity
between the market status in the past and that at present. For example, Gyorfi et al.~\cite{gyorfi2006nonparametric,gyorfi2008nonparametric}
uses Euclidean distance, while Li et al.~\cite{li2011corn} adopts the Pearson correlation coefficient.
Empirical studies demonstrate that the correlation-based pattern matching approach, denoted by CORN-K,
can generally achieve better performance than other pattern matching-based methods~\cite{li2011corn}.



In spite of the success of CORN-K (and some other pattern matching methods), a potential problem of this approach
is that no risk is considered when searching for optimal portfolios, i.e., the second stage of the algorithm.
This is clearly a shortcoming as risky portfolios will lead to reduced long-term return.
This problem is particularly severe for volatile markets that involve many risky assets.
A natural idea is to penalize the risky portfolios when searching for the optimal portfolio.
In this work, we propose a risk-aversion CORN-K algorithm, RACORN-K, that penalizes
risky portfolios by adding a regularization term in the optimization objective function.
We evaluate this new algorithm with four datasets (DJIA, MSCI, SP500(N), HSI).
The results demonstrate that RACORN-K delivers notable and consistent performance improvements,
in terms of long-term return, Sharp ratio and maximum drawdown. The improvements
on the volatile markets DJIA and SP500(N) are particularly remarkable, demonstrating the
value of the proposal.

\section{Problem Setting}
\label{sec:proset}

Consider an investment over $m$ assets on $n$ trading periods. Define the relative price vector at trading period $t$ by $\mathbf{x}_t=(x_{t,1},...,x_{t,m}) \in \mathbb{R}^m_{+}$, whose $i$-th component $x_{t,i}=\frac{P(t,i)}{P(t-1,i)}$ and $P(t,i)$ is the closing price of the $i$-th asset at the $t$-th trading period. Given a window size
$w$, the market window for period $t$ is defined as $\mathbf{X}^{t-1}_{t-w}=(\mathbf{x}_{t-w},...,\mathbf{x}_{t-1})$, which is assumed to represent the status of the market at period $t$.

A portfolio denoted by $\mathbf{b}_t = (b_{t,1},...,b_{t,m})^T \in \mathbb{R}^m$ is defined as a distribution over the $m$ assets, where $b_{t,i}$ is the proportion of the investment on the $i$-th asset at period $t$. In this study, we assume that only long positions are allowed, which implies
the following constraint on $\mathbf{b}_t$: $b_{t,i}\ge0, \sum_ib_{t,i}=1$.

At the trading period $t$, an investor selects a portfolio $\mathbf{b}_t$ given the past market relative prices $\{\mathbf{x}_1,...,\mathbf{x}_{t-1}\}$. The instant return is computed by $s_t = \sum_i b_{t,i}x_{t,i} = \mathbf{b}_t^T \mathbf{x}_t$, and the accumulated return produced by $ \{\mathbf{b}_1,...,\mathbf{b}_n \}$ is $S_t = \prod_{j=1}^{n} \mathbf{b}_j^T \mathbf{x}_j $.

\section{Algorithm}
\label{sec:algo}

In this section, we first give a brief description of the classical CORN-K algorithm, and then propose
our RACORN-K algorithm. A conservative version of RACORN-K, denoted by RACORN(C)-K, will be also proposed.

\subsection{CORN-K algorithm}
\label{sec:corn}

At the $t$-th trading period, the CORN-K algorithm first selects all the historical periods whose
market status is similar to that of the present market, where the similarity is measured by the Pearson correlation
coefficient. This patten matching process produces a set of similar periods, denoted by $C(\mathbf{x}_t; w, \rho)$,
where $w$ is the size of the market window, and $\rho$ is the threshold when selecting similar periods. This is
formulated as follows:

\[
C(\mathbf{x}_t; w, \rho) = \{ \mathbf{x}_j | corr(\mathbf{X}^{j-1}_{j-w}, \mathbf{X}^{t-1}_{t-w})> \rho \},
\]
\noindent where $corr(X,Y)$ is the correlation coefficient between $X$ and $Y$, and $w < j < t$.
Note that when calculating the correlation coefficient, the columns in $\mathbf{X}^{j-1}_{j-w}$ (the same for 
$\mathbf{X}^{t-1}_{t-w}$) are concatenated into a $(m \times w)$-dimensional vector.
Once the similar periods have been selected, the portfolio on the $m$ assets can be obtained following the
BCRP principle~\cite{cover1986empirical}:

\begin{equation}
\label{eq:opt}
\mathbf{b}^*_t(w,\rho) = \mathop{argmax}_{\mathbf{b} \in \Delta_m} \sum_{\mathbf{x} \in C(\mathbf{x}_t; w,\rho)} log( \mathbf{b}^T \mathbf{x}),
\end{equation}

\noindent where $\Delta_m = \{\mathbf{b}: \sum_{i=1}^{m} b_i = 1, b_i \ge 0\}$ represents a simplex with $m$ components.

Finally, CORN-K selects various $w$ and $\rho$. By each setting of these parameters $(w,\rho)$, an optimal
portfolio $\mathbf{b}^*_t(w,\rho)$ is computed following (\ref{eq:opt}). Note that $\mathbf{b}^*_t(w,\rho)$ is a particular strategy,
also called an `expert', denoted by $\epsilon(w,\rho)$.
The experts which achieve top-k accumulated returns are selected to compose an expert ensemble $E_t$,
where the accumulated return of an expert $\epsilon(w,\rho)$ is denoted by $S_{t}(w, \rho)$. With the expert ensemble $E_t$,
the ensemble-based optimal portfolio is derived by:

\begin{equation}
\label{eq:ens}
\mathbf{b}^*_t = \frac{\sum_{\epsilon(w, \rho) \in E_t } S_{t-1}(w, \rho) \mathbf{b}^*_t (w, \rho)}{\sum_{\epsilon(w, \rho) \in E_t } S_{t-1}(w, \rho)}.
\end{equation}

\noindent It is expected that this ensemble-based average leads to more robust portfolios.

\subsection{Risk-Aversion CORN-K (RACORN-K)}
\label{sec:ocr}

The portfolio optimization is crucial for the success of CORN-K. A potential problem of the existing
form~(\ref{eq:opt}), however, is that the optimization is purely profit-driven.
This is clearly dangerous as the high-profit assets it selects may exhibit 
high variation, leading to a risky portfolio that suffers from unexpected loss.
It is particularly true for volatile markets where the prices of many assets
are unstable.
A natural idea to solve this problem is to penalize risky portfolios when searching
for the optimal portfolio. This leads to a risk-aversion CORN-K, denoted by RACORN-K.
More specifically, the objective function in (\ref{eq:opt}) is augmented by
a risk-penalty term, formulated as follows:

\begin{equation}
\label{eq:risk}
\mathbf{b}^*_t (w, \rho, \lambda)= \mathop{argmax}_{\mathbf{b} \in \Delta_m}  \frac{\sum_{\mathbf{x} \in C(\mathbf{x}_t; w, \rho)} log( \mathbf{b}^T \mathbf{x})}{|C(\mathbf{x}_t; w, \rho)|} -\lambda \sigma_t(w,\rho)
\end{equation}

\noindent where $\lambda$ is the risk-aversion coefficient, $|C(\mathbf{x}_t; w, \rho)|$ is the size of $C(\mathbf{x}_t; w, \rho)$ ,
and $\sigma_t(w,\rho)$ is the risk:

\[
\sigma_t(w,\rho) = std(log( \mathbf{b}^T \mathbf{x}))|_{\mathbf{x} \in C(\mathbf{x}_t; w,\rho)},
\]
where $std(\cdot)$ denotes the standard deviation function.

We emphasize that the risk-penalty term $std( log(\mathbf{b}^T \mathbf{x}))$
is different from $\mathbf{b}^T std(log(\mathbf{x}))$: the former is the risk of the
\emph{portfolio}, while the latter is the sum of the risk of the assets \emph{according to the portfolio}.
This form is similar to the classical mean-variance model~\cite{Markowitz1991Portfolio}.
A key difference from the mean-variance model (and most other risk-aversion models) is that the risk is computed
over the historical price relatives in $C(\mathbf{x}_t; w,\rho)$, rather than on the whole trading periods.
It therefore estimates the risk of the portfolio with a particular pattern matching strategy, i.e., the CORNK-K algorithm, 
rather than the unconstrained market risk of the selected assets.

With the new optimization objective (\ref{eq:risk}), the ensemble-based optimal portfolio is derived similarly as in
CORN-K. The only difference is that we have introduced a new hyper-parameter $\lambda$, so the expert should
be extended to $\epsilon(w, \rho, \lambda)$. The derivation is similar to (\ref{eq:ens}), formulated by:

\[
\mathbf{b}^*_t = \frac{\sum_{\epsilon(w, \rho, \lambda) \in E_t } S_{t-1}(w, \rho, \lambda) \mathbf{b}^*_t (w, \rho, \lambda)}{\sum_{\epsilon(w, \rho, \lambda) \in E_t } S_{t-1}(w, \rho, \lambda)}.
\]

%


\subsection{Conservative RACORN-K (RACORN(C)-K)}

In the above RACORN-K algorithm, the risk-aversion coefficient $\lambda$ is treated as a new free parameter
and is combined with $w$ and $\rho$ to derive ensemble-based optimal portfolio. A potential problem of this
type of `naive ensemble' is that it does not consider the time-variant property of the risk. In fact,
the risk of the portfolio derived from each expert tends to change quickly in an volatile market
and therefore the weights of individual experts should be adjusted timely.
To achieve the quick adjustment, we use the instant return $s_t(w,\rho, \lambda)$ to weight
the experts with different $\lambda$, rather than the accumulated return $S_{t-1}(w,\rho,\lambda)$.
This is formulated as follows:

\begin{equation}
\label{eq:h1}
\mathbf{b}^*_t(w, \rho) = \frac{\sum_{\lambda} s_{t}(w, \rho, \lambda) \mathbf{b}^*_t (w, \rho, \lambda)}{\sum_{\lambda} s_{t}(w, \rho, \lambda)}.
\end{equation}

\noindent Since $s_t$ is not available when estimating $\mathbf{b}^*_t $,
we approximate it by the geometric average of the returns achieved in $C(\mathbf{x}_t; w,\rho)$, given by:

\[
s_{t}(w, \rho, \lambda) \approx exp(\frac{\sum_{\mathbf{x}_j \in C(\mathbf{x}_t; w,\rho)} log(\mathbf{b}_t^*(w,\rho,\lambda)^T \mathbf{x}_j)}{|C(\mathbf{x}_t; w, \rho)|}).
\]
\noindent
In practice, we find that omitting the normalization term $\frac{1}{|C(\mathbf{x}_t; w,\rho)|}$ 
can deliver slightly better results.

Once $\mathbf{b}^*_t(w, \rho)$ is obtained, the ensemble-based optimal portfolio can be derived as
CORN-K following (\ref{eq:ens}), where
the accumulated return $S_{t-1}(w,\rho)$ is achieved by applying the portfolios derived by (\ref{eq:h1}).
Compared to RACORN-K, this variant algorithm is more risk-aware and thus assumed to be more conservative.
We denote this conservative version of RACORN-K as RACORN(C)-K.

\section{Experiments}
\label{sec:exp}

We evaluate RACORN-K and RACORN(C)-K on four datasets, and compare the performance with the classical CORN-K algorithm.
The performances of some other popular strategies are also reported.

\subsection{Dataset}
\label{sec:dataset}

\begin{table}[!hbpt]
\scriptsize
\caption{Datasets used in the experiments. }
\label{tab:dataset}
\centering
\begin{tabular}{ccccc}
\toprule
Dataset & Region & Time range & Trading days & Assets\\
\midrule
DJIA & US & 2001/01/14 - 2003/01/14 & 507 & 30\\
MSCI & GLOBAL & 2006/04/01 - 2010/03/31 & 1043 & 24\\
SP500(N) & US & 2000/01/03 - 2014/12/31 & 3773 & 10\\
HSI & HK & 2000/01/03 - 2014/12/31 & 3702 & 10\\
\bottomrule
\end{tabular}
\end{table}

Table~\ref{tab:dataset} shows the four datasets used in our experiments.
The DJIA (Dow Jones Industrial Average) dataset is a collection of $30$ large publicly
owned companies based in the United States, collected by Borodin et al.~\cite{borodin2004can}.
The MSCI\footnote{http://olps.stevenhoi.org/} dataset is a collection of $24$ global equity indices which are the constituents
of MSCI World Index.
These two datasets are relatively old. In order to evaluate the performance of the
proposed algorithms on more recent data, we collected another two datasets:
SP500(N) and HSI. The SP500(N) dataset consists of $10$ equities with the largest market capitalization
(as of Apr. 2003) from the S\&P 500 Index. Note that this dataset is different from
the SP500 dataset collected by Li et al.~\cite{li2011corn}. The latter is a little old
and may not reflect the trend of the current market\footnote{In fact,
our new method also performs well on the old SP500 dataset. See the extended version of this
paper (http://project.cslt.org).}.
The HSI dataset contains $10$ equities with the largest market capitalization (as of Jan. 2005)
from Hang Seng Index. It is worth noting that SP500(N) and HSI cover both bull markets and bear markets,
particularly the finance crisis in 2009.

\subsection{Implement details}

The OLPS toolbox~\cite{li2015olps} is used to implement the baseline strategies, where
the default values are set for the hyper-parameters. For RACORN-K, the maximum window size is set to $5$.
The correlation coefficient threshold $\rho$ ranges from $0$ to $0.9$, with the step set to be $0.1$.
The risk-aversion coefficient $\lambda$ ranges from $0$ to $0.03$, with a step $0.01$. While combining the outputs of experts,
top 10\% experts are selected to compose the ensemble $E_t$.
As for RACORN(C)-K, all the parameters are the same as RACORN-K, except that $\lambda$
ranges from $0$ to $0.1$ with a step $0.01$, as we found RACORN(C)-K has the capability to accept a
larger maximum risk aversion. These parameters are used in the experiments on all the four datasets.

\subsection{Experimental results}

Three metrics are adopted to evaluate the performance of a strategy: accumulated return (RET),
Sharpe ratio (SR) and maximum drawdown (MDD). Among these metrics, SR and MDD are more
concerned as our main goal is to control the risk.
And the improvement on SR and the reduction on MDD are often
more important for investors, particularly for asset managers who can leverage various financial tools to magnify returns.

\begin{table*}[!hbpt]
\caption{Performances of different strategies on four datasets.}
\label{tab:results}
\footnotesize
\centering
\scalebox{0.8}{
\begin{tabular}{c|cccc|ccc|ccc|ccc}
\toprule
&Dataset & \multicolumn{3}{c}{DJIA}  &  \multicolumn{3}{c}{MSCI} & \multicolumn{3}{c}{SP500(N)} & \multicolumn{3}{c}{HSI}  \\
\midrule
&Criteria  & RET  & SR  & MDD  & RET  & SR  & MDD  & RET  & SR  & MDD  & RET  & SR  & MDD   \\
\midrule
\multirow{3}{2cm}{Main Results}
&RACORN(C)-K			&	\textbf{0.93}  &   \textbf{0.01}  & \textbf{0.32}  &	\textbf{78.38}	&   \textbf{3.73}	&   \textbf{0.21} 	&	\textbf{12.55}  & \textbf{0.77}  & \textbf{0.53} &	202.04  &   \textbf{1.60}  & \textbf{0.28}  \\

&RACORN-K&	\textbf{0.83}  &   \textbf{-0.19} &   \textbf{0.37}  &	\textbf{79.52}	&   \textbf{3.67}   &   \textbf{0.21} 	&	\textbf{13.03} & \textbf{0.72} & \textbf{0.57}&	\textbf{264.02} &  \textbf{1.60} &   \textbf{0.29} \\

&CORN-K									&   0.80   & -0.24  &   0.38  &	77.54 			& 	3.63   &   0.21 &	12.50  &   0.70 &   0.60&	254.27  &   1.56 &  0.30 \\

\midrule
\midrule
\multirow{2}{2cm}{Naive Methods}
&UBAH			&	0.76	&	-0.43	&	0.39 &	0.90	&	0.02	&	0.65 &	1.52	&	0.24	&	0.50 &	3.54	&	0.53	&	0.58 \\
&UCRP			&	0.81	&	-0.28	&	0.38 &	0.92	&	0.05	&	0.64 &	1.78	&	0.28	&	0.68 &	4.25	&	0.58	&	0.55 \\
\midrule
\multirow{3}{2cm}{Follow the Winner}
&UP				&	0.81	&	-0.29	&	0.38 &	0.92	&	0.04	&	0.64 &	1.79	&	0.29	&	0.68 &	4.26	&	0.59	&	0.55 \\
&EG				&	0.81	&	-0.29	&	0.38 &	0.92	&	0.04	&	0.64 &	1.75	&	0.28	&	0.67 &	4.22	&	0.58	&	0.55 \\
&ONS			&	1.53	&	0.80	&	0.32 &	0.85	&	0.02	&	0.68 &	0.78	&	0.27	&	0.96 &	4.42	&	0.52	&	0.68 \\
\midrule
\multirow{6}{2cm}{Follow the Loser}
&ANTICOR		&	1.62	&	0.85	&	0.34 &	2.75	&	0.96	&	0.51 &	1.16	&	0.24	&	0.93 &	9.10	&	0.74	&	0.56 \\
&ANTICOR2		&	2.28	&	1.24	&	0.35 &	3.20	&	1.02	&	0.48 &	0.71	&	0.22	&	0.97 &	12.27	&	0.77	&	0.55 \\
&PAMR\_2		&	0.70	&	-0.15	&	0.76 &	16.73	&	2.07	&	0.54 &	0.01	&	-0.28	&	1.00 &	1.19	&	0.20	&	0.86 \\
&CWMR\_Stdev&	0.69	&	-0.17	&	0.76 &	17.14	&	2.07	&	0.54 &	0.02	&	-0.26	&	0.99 &	1.28	&	0.22	&	0.85 \\
&OLMAR1		&	2.53	&	1.16	&	0.37 &	14.82	&	1.85	&	0.48 &	0.03	&	-0.11	&	1.00 &	4.19	&	0.46	&	0.77 \\
&OLMAR2		&	1.16	&	0.40	&	0.58 &	22.34	&	2.08	&	0.42 &	0.03	&	-0.11	&	1.00 &	3.65	&	0.43	&	0.84 \\
\midrule
\multirow{2}{2cm}{Pattern Matching based Algorithms}
&BK				&	0.69	&	-0.68	&	0.43 &	2.62	&	1.06	&	0.51 &	1.97	&	0.31	&	0.59 &	13.90	&	0.88	&	0.45 \\
&BNN			&	0.88	&	-0.15	&	0.31 &	13.40	&	2.33	&	0.33 &	6.81&	0.67	&	0.41 &	104.97&	1.40	&	0.33 \\
\bottomrule
\end{tabular}
}
\end{table*}

\begin{figure}[htb]
\centering
\subfigure[DJIA]{
\begin{minipage}[b]{0.22\textwidth}
\includegraphics[width=1\textwidth]{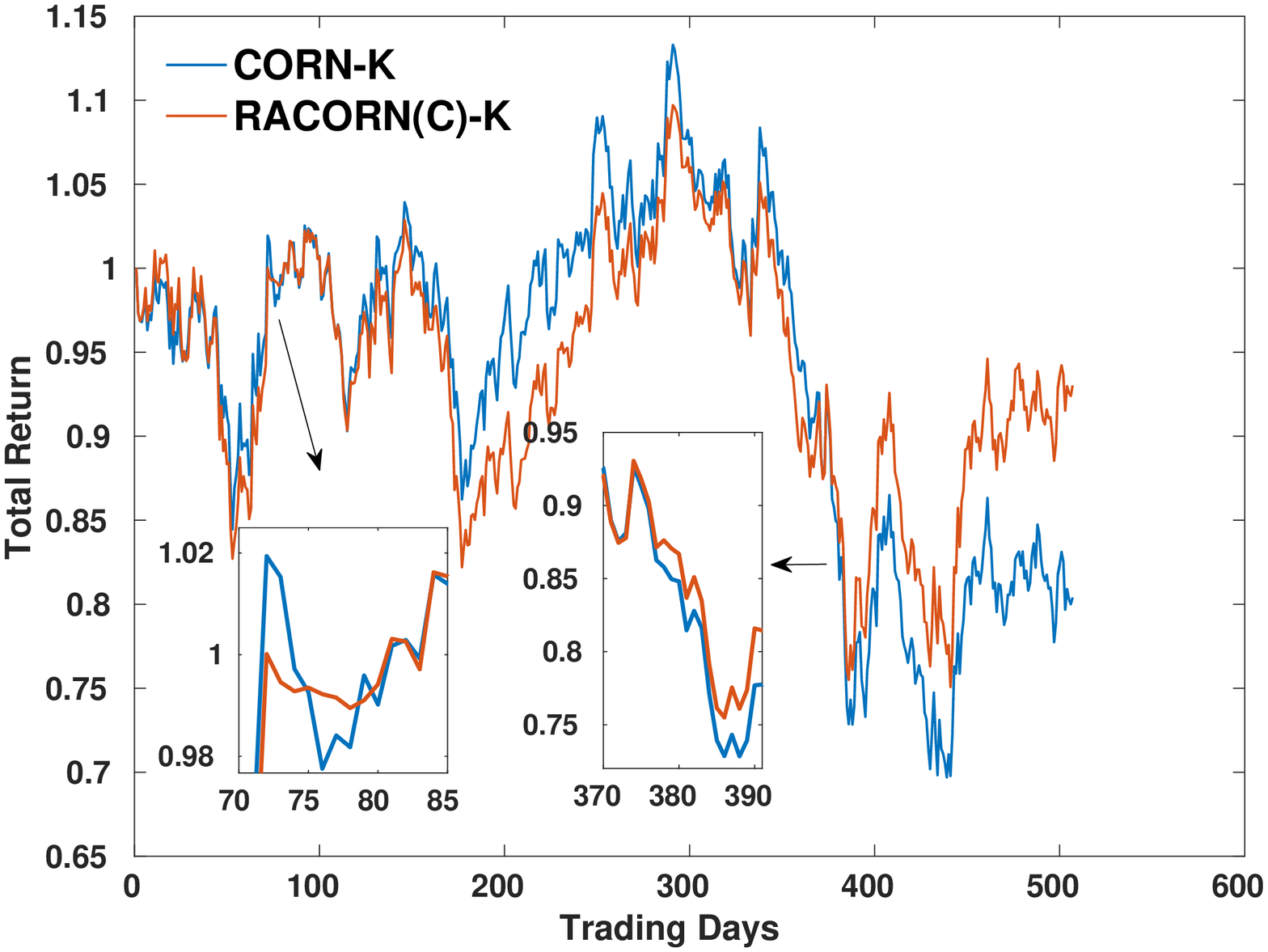}
\end{minipage}
}
\subfigure[MSCI]{
\begin{minipage}[b]{0.22\textwidth}
\includegraphics[width=1\textwidth]{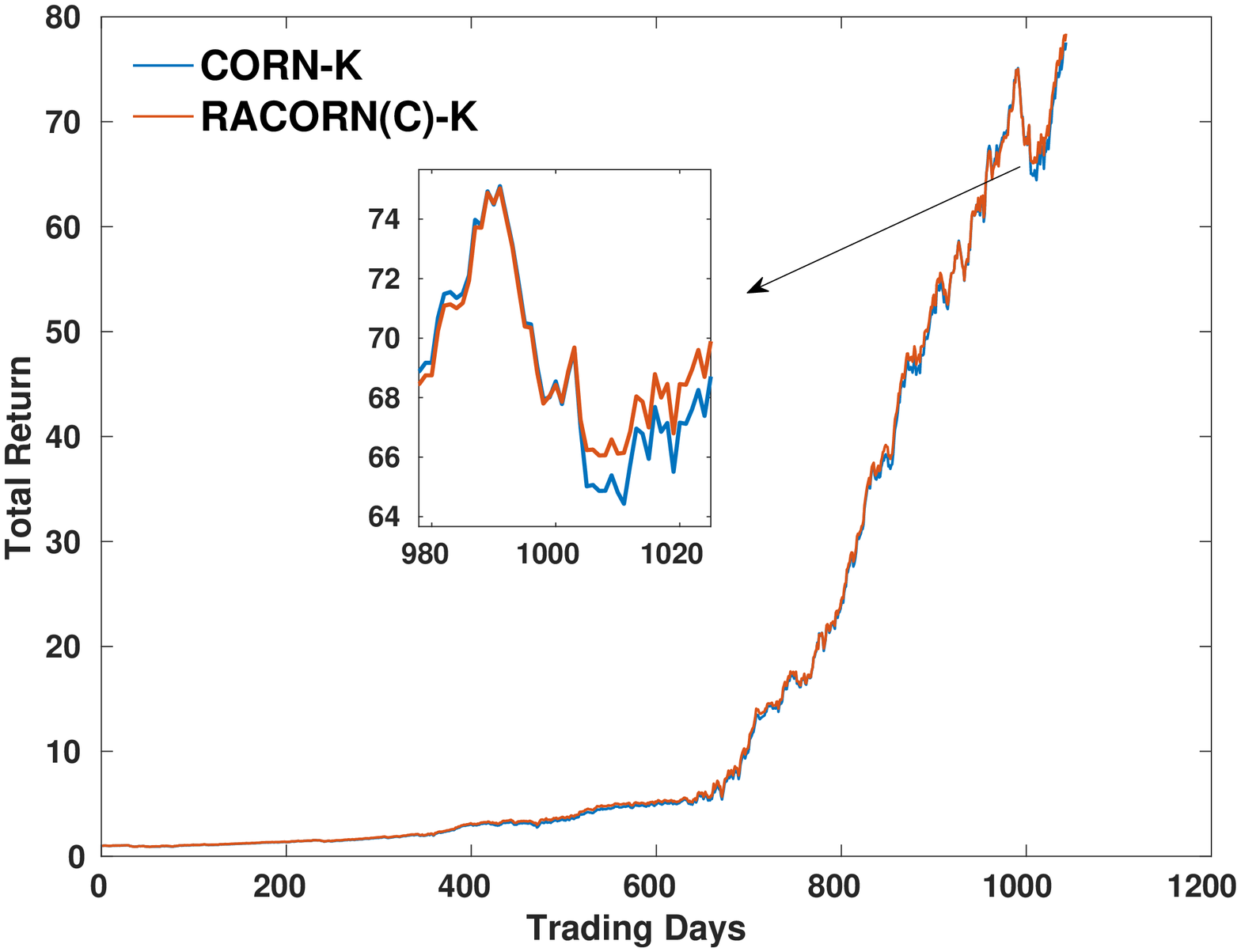}
\end{minipage}
}

\subfigure[SP500(N)]{
\begin{minipage}[b]{0.22\textwidth}
\includegraphics[width=1\textwidth]{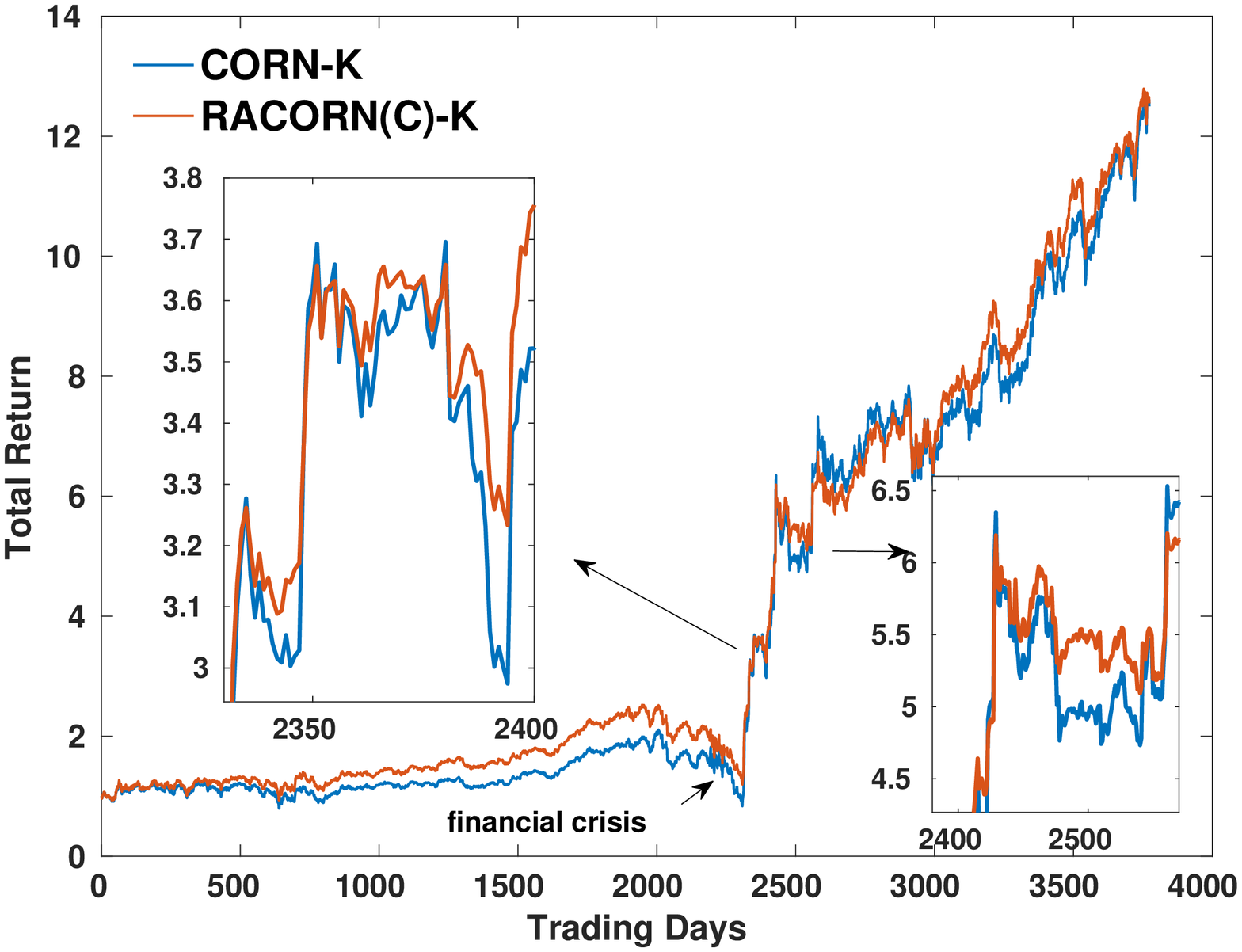}
\end{minipage}
}
\subfigure[HSI]{
\begin{minipage}[b]{0.22\textwidth}
\includegraphics[width=1\textwidth]{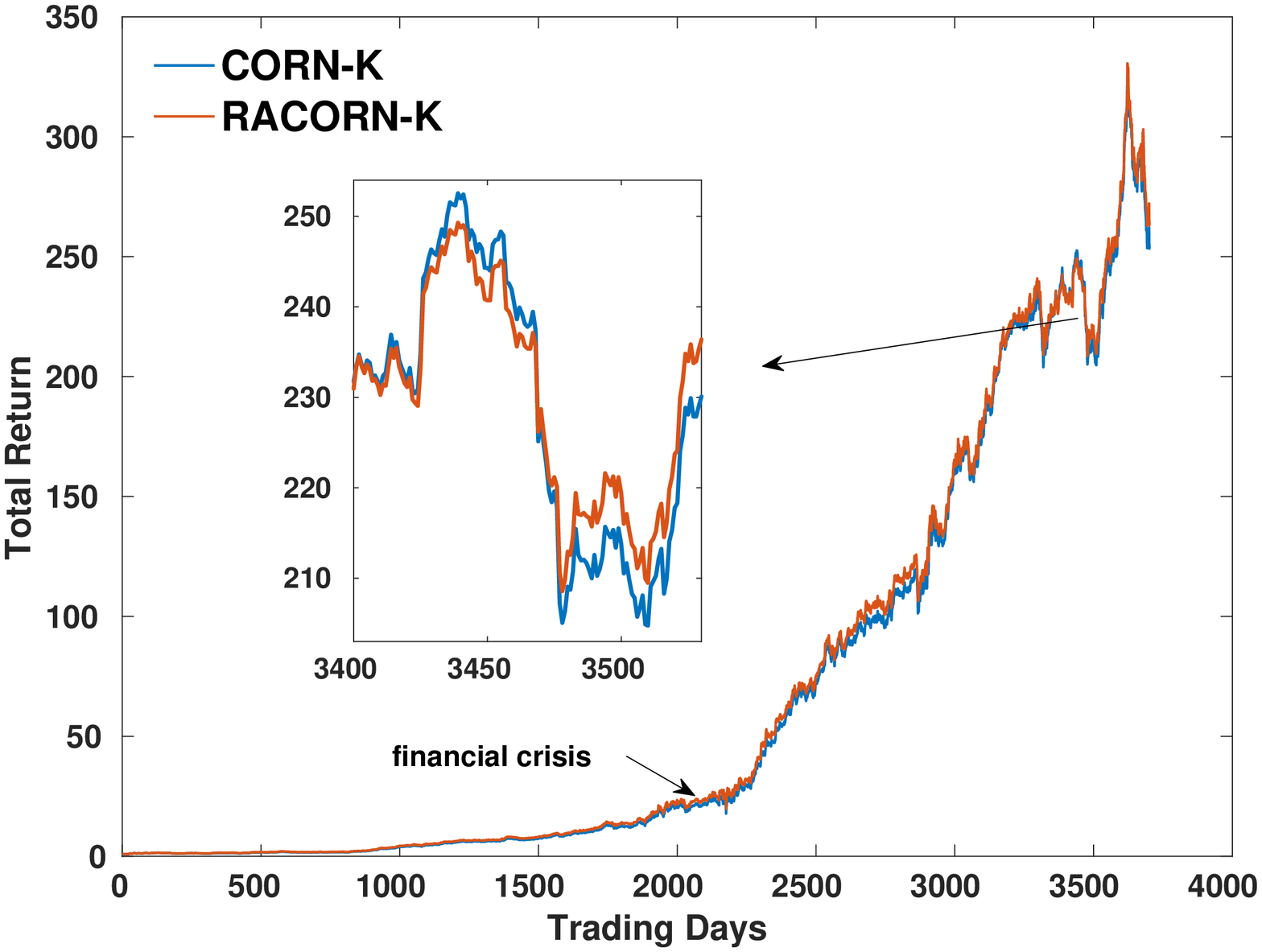}
\end{minipage}
}
\caption{RET curves with CORN-K and RACORN-K/RACORN(C)-K on four datasets.}
\label{fig:figrst}
\end{figure}

\subsubsection{General results}

Table~\ref{tab:results} summarizes the results, where the improvements compared
to CORN-K have been marked as bold face.
From these results, it can be seen that RACORN-K consistently improves SR
and MDD on all the datasets, which confirms that involving risk aversion does
reduce the risk of the derived portfolio. The conservative version, RACORN(C)-K, delivers
even better performance in terms of SR and MDD, though the long-term
return (RET) is slightly reduced.
In most cases, both RACORN-K and RACORN(C)-K obtain larger RETs than the CORN-K baseline,
demonstrating that controlling risk will ultimately improve long-term profits. The only
exception is that the RET on HSI drops with RACORN(C)-K; however, the absolute RET has been very
high, so this RET reduction can be regarded as a reasonable cost for the risk control.

Fig.~\ref{fig:figrst} shows the RET curves with CORN-K and RACORN-K/RACORN(C)-K.\footnote{RACORN-K 
rather than RACORN(C)-K is plotted for HSI as its RET curve better matches the RET curve of CORN-K, 
so readers can see more clearly how the risk-aversion penalty changes the behavior of the algorithm.}
It can be seen that
RACORN-K or RACORN(C)-K has the same trend as CORN-K in general, particularly on relatively stable
markets (MSCI and HSI). However, in periods where CORN-K is risky, RACORN(C)-K behaves
less bumpy and hence more reliable. This can be seen clearly in (a) DJIA and (c) SP500(N).
Fig.~\ref{fig:figrst} presents some `key points' where
RACORN(C)-K behaves more `smooth' than CORN-K. Due to this smoothness, the risk of the
strategy is reduced, and extremely poor trading can be largely avoided.

When comparing to other baselines, it can be seen that the CORN-K family performs much
better and more consistent. For example, OLMAR1, a classical follow-the-loser
strategy, performs the best on DJIA, but the advantage is totally lost on other datasets.
These results re-confirm the reliability of pattern matching methods.

\subsubsection{Detailed analysis}

Analyzing the performance of RACORN-K/RACORN(C)-K on different markets sheds more light on
the property of the risk-aversion approach.
From Table~\ref{tab:results}, we can see that RACORN-K/RACORN(C)-K obtains the most significant
SR improvement on DJIA, and the most significant MDD reduction on DJIA and SP500(N).
Interestingly, these two datasets are the ones that the conventional CORN-K does not work
well (less RET, smaller SR, higher MDD). This can be also seen from Fig.~\ref{fig:figrst},
where the RET curves with CORN-K exhibit more risk on DJIA and SP500(N) compared to
the curves on MSCI and HSI. This indicates that RACORN-K/RACORN(C)-K are more effective
when the conventional CORN-K is risky.

\begin{figure}[htb]
\centering
\subfigure[DJIA]{
\begin{minipage}[b]{0.22\textwidth}
\includegraphics[width=1\textwidth]{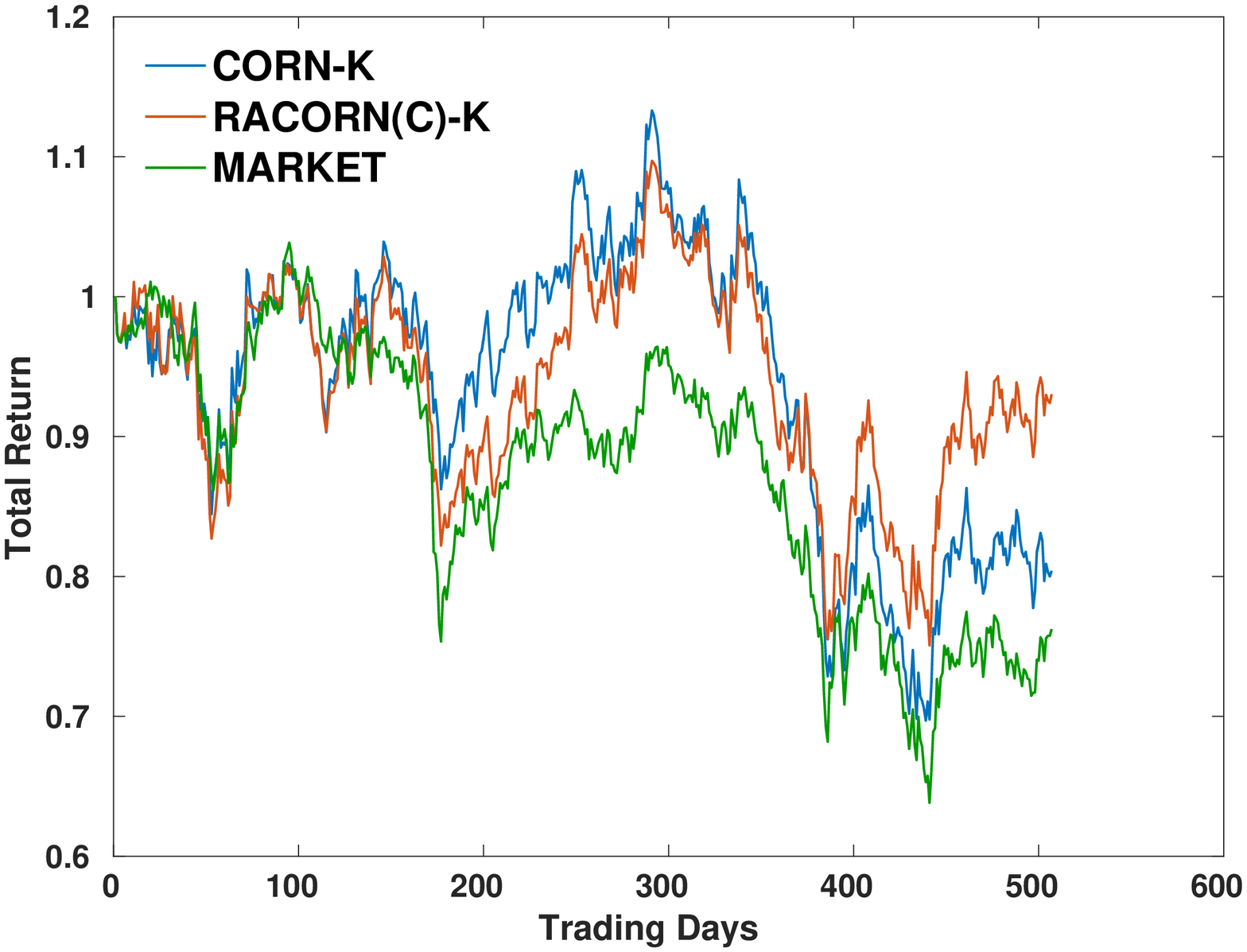}
\end{minipage}
}
\subfigure[SP500(N)]{
\begin{minipage}[b]{0.22\textwidth}
\includegraphics[width=1\textwidth]{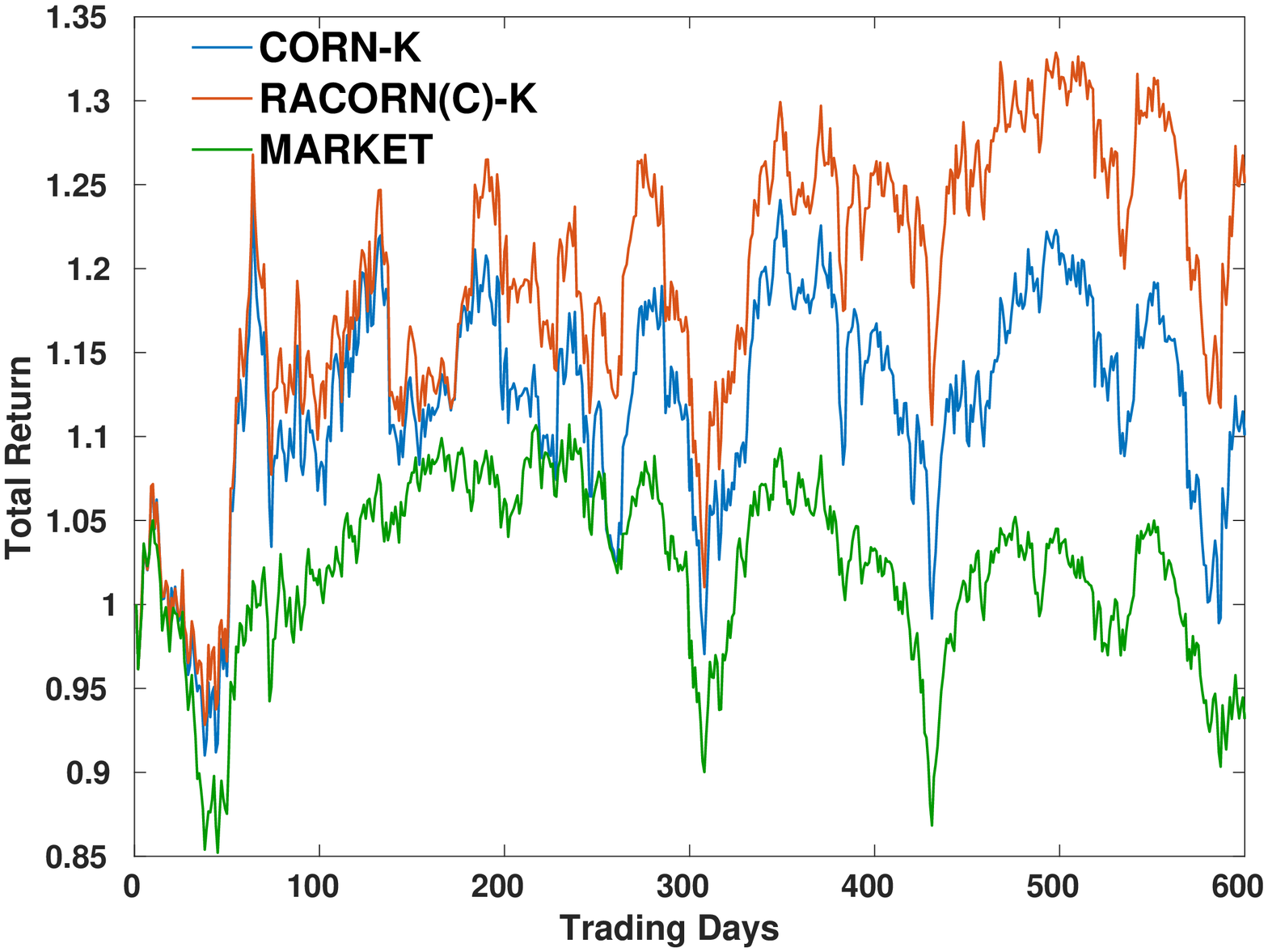}
\end{minipage}
}
\caption{Market and RET curves with CORN-K and RACORN(C)-K on DJIA and SP500(N).}
\label{fig:mkt}
\end{figure}

More analysis shows that the risk of CORN-K is largely attributed to the risk of the market.
To make it clear, the market returns of DJIA and SP500(N) are plotted together with the RET
curves of CORN-K and RACORN(C)-K in Fig.~\ref{fig:mkt}. For a clear presentation, only the first 600 trading
days of SP500(N) are plotted as during this period the market is volatile.
It shows clearly that
on the markets with huge volatility, involving risk-aversion largely reduced
the risk, hence a more reliable strategy. As a summary, CORN-K may perform less effective on
risky markets, and the risk-aversion algorithms can largely alleviate this problem.
Fortunately, this advantage on risky markets does not degrade its performance
on stable markets (where CORN-K works well). This is a nice property and indicates
that RACORN-K/RACORN(C)-K is a safe and effective extension/substitution of CORN-K.



\section{Conclusion}
\label{sec:conclusion}

This paper presented two risk-aversion CORN-K algorithms, RACORN-K and RACORN(C)-K
that involve a risk-aversion penalty when searching for optimal portfolios.
Experimental results on four datasets
demonstrate that the new algorithms can consistently improve the Sharpe ratio and
reduce maximum drawdown. This improvement is particularly significant on
high-risk markets where the conventional CORN-K tends to perform not as well.
Fortunately, this risk control does not degrade the long-term profit in general, and
in many cases, it leads to even better returns. Future work involves exploring 
more suitable regularizations, e.g., group penalty and temporal continuity constraint.

\vfill\pagebreak

%

\bibliographystyle{IEEEbib}
\bibliography{refs}

\end{document}